\documentclass[conference, letterpaper]{IEEEtran}
\usepackage{float}
\usepackage{amssymb}
\usepackage{amsmath}
\usepackage[letterpaper, top=1.92cm, bottom=4cm, left=1.92cm, right=1.92cm]{geometry}
\usepackage{spconf,amsmath,graphicx}
\usepackage{enumitem}
\setlist{nosep, leftmargin=14pt}
\usepackage{cite}
\usepackage[backref]{hyperref}
\hypersetup{hidelinks,
	colorlinks=true,
	allcolors=black,
	pdfstartview=Fit,
	breaklinks=true}

\usepackage{mwe} 
\usepackage{multirow}
\usepackage{booktabs}

\title{AI-Driven Automated Tool for Abdominal CT Body Composition Analysis in Gastrointestinal Cancer Management}
\name{Xinyu Nan$^{1}$ \qquad Meng He$^{2}$ \qquad Zifan Chen$^{1}$ \qquad Bin Dong$^{3,4,5\star}$ \qquad Lei Tang$^{2\star}$ \qquad Li Zhang$^{1\star}$}
\address{$^{1}$Center for Data Science, Peking University, China\\$^{2}$Department of Radiology, Key Laboratory of Carcinogenesis and Translational Research(Ministry of \\Education), Peking University Cancer Hospital and Institute, Beijing, China\\$^{3}$Beijing International Center for Mathematical Research (BICMR), Peking University, Beijing, China\\$^{4}$Center for Machine Learning Research, Peking University, Beijing, China\\$^{5}$National Biomedical Imaging Center, Peking University, Beijing, China}

\begin{document}
%
\maketitle

\vspace{-0.5cm}
\begin{abstract}
The incidence of gastrointestinal cancers remains significantly high, particularly in China, emphasizing the importance of accurate prognostic assessments and effective treatment strategies. Research shows a strong correlation between abdominal muscle and fat tissue composition and patient outcomes. However, existing manual methods for analyzing abdominal tissue composition are time-consuming and costly, limiting clinical research scalability. To address these challenges, we developed an AI-driven tool for automated analysis of abdominal CT scans to effectively identify and segment muscle, subcutaneous fat, and visceral fat. Our tool integrates a multi-view localization model and a high-precision 2D nnUNet-based segmentation model, demonstrating a localization accuracy of 90\% and a Dice Score Coefficient of 0.967 for segmentation. Furthermore, it features an interactive interface that allows clinicians to refine the segmentation results, ensuring high-quality outcomes effectively. Our tool offers a standardized method for effectively extracting critical abdominal tissues, potentially enhancing the management and treatment for gastrointestinal cancers.
The code is available at \href{https://github.com/NanXinyu/AI-Tool4Abdominal-Seg.git}{https://github.com/NanXinyu/AI-Tool4Abdominal-Seg.git}.
\end{abstract}
%
\section{Introduction}

Gastrointestinal cancers pose a significant global health challenge, with approximately 1.2 million new cases diagnosed annually worldwide, 40\% of which occur in China~\cite{joshi2021current}. This high incidence underscores the critical need for effective evaluation of patient prognosis and treatment efficacy. Recent studies have revealed a strong association between abdominal body composition and patient outcomes in gastrointestinal cancers~\cite{He2023AssociationsOS,kim2021prognostic,chen2023sarcopenia,xu2019prognostic}. Sarcopenia, for instance, has been linked to poor prognosis, particularly in patients receiving targeted therapies or immune checkpoint inhibitor (ICI) treatments~\cite{kim2021prognostic,chen2023sarcopenia}. Conversely, the "obesity paradox" suggests that increased subcutaneous and visceral fat may serve as a protective factor, especially for patients undergoing ICI therapies~\cite{xu2019prognostic}.

Given the significant impact of abdominal tissue composition on patient outcomes, accurate measurement and analysis of these tissues are crucial. However, current clinical methods primarily rely on manual annotation by specialists, a process that is time-consuming, costly, and limits large-scale clinical research~\cite{koitka2021fully,park2020development}. This limitation highlights the urgent need for an automated tool to streamline body composition assessment, enabling large-scale analysis and potentially improving patient care through more accurate and efficient evaluation.

Automated analysis of abdominal tissue composition relies on accurate region-of-interest (ROI) localization and precise tissue segmentation. Recent advancements in artificial intelligence (AI) have significantly improved both areas. Chen et al. demonstrated AI's potential in anatomical region identification with their multi-scale context-guided model for spinal tissue localization~\cite{chen2022multi}. In segmentation, Ronneberger et al. introduced the U-Net architecture~\cite{ronneberger2015u}, later expanded into nnU-Net~\cite{nnunet2021}, which has shown robust performance across various clinical tasks. Jothiraj et al. also proposed a deep learning 
 framework~\cite{10053299} consisting of localization and semantic segmentation of polyp for early diagnosis of colorectal cancer.
\begin{figure*}[htbp]
\centering
\includegraphics[width=0.75\textwidth]{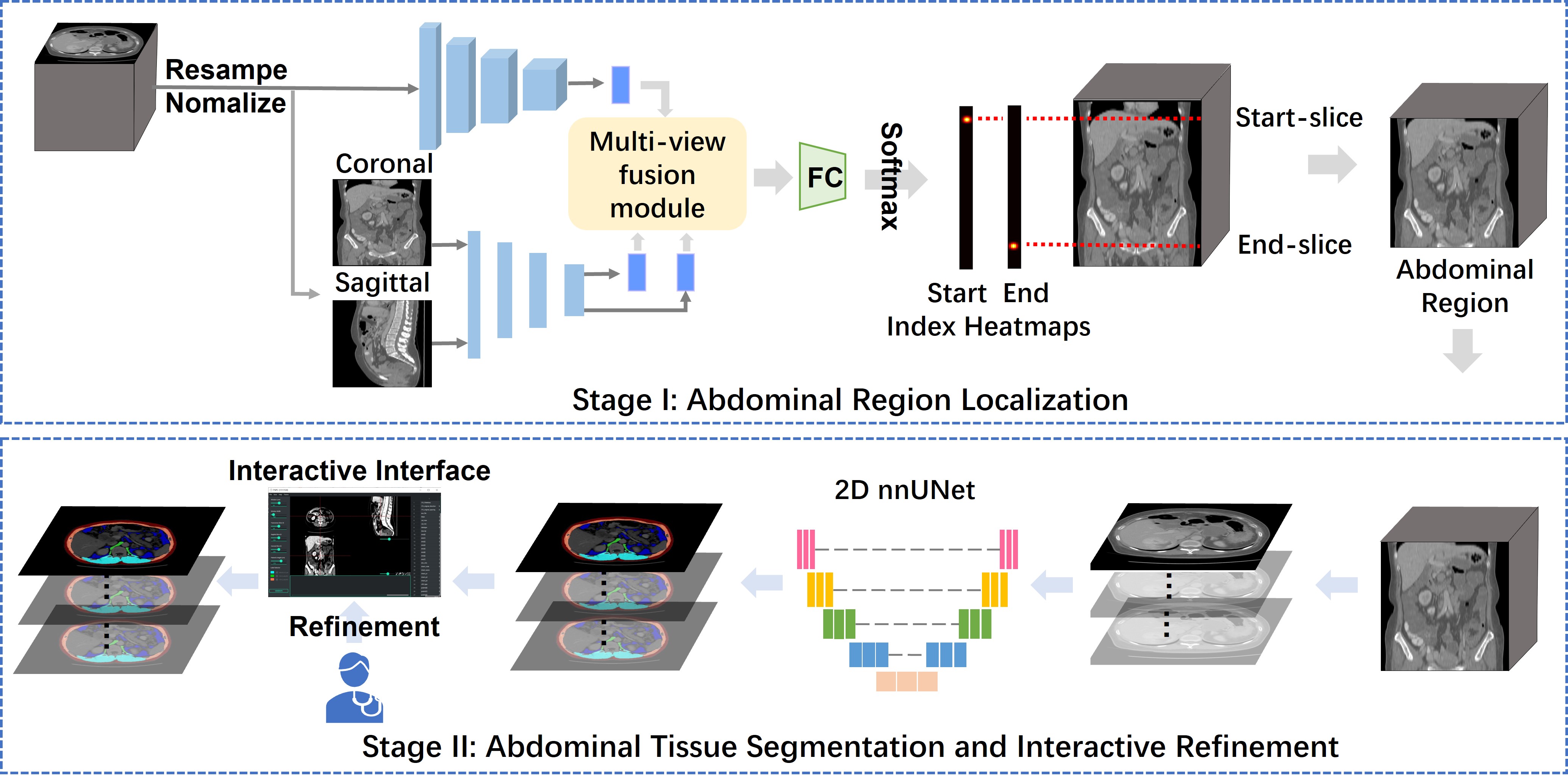}
\caption{An overview of our automatic annotation tool for abdominal components. (a) Stage I: A multi-view localization model consisting of a 3D ResNet18, a lightweight downsampling network and a multi-view fusion module to localize the start and end slice indexes of abdominal slices. (b) Stage II: A two-step process involving (i) a high-precision segmentation model based on 2D nnUNet architecture to segment abdominal muscle, SFA, and VFA of each slice, producing high-quality initial results, and (ii) an optional interactive interface for clinicians to further refine these segmentation results if needed.}
\label{fig:overview}
\end{figure*}
Leveraging these advances, we developed an automated tool for efficient abdominal tissue analysis in gastrointestinal cancer management. Our proposed tool integrates two key components: (1) A multi-view abdominal localization model, which accurately detects the starting and ending slice positions of the abdomen within a CT scan using a novel multi-view fusion module. (2) A comprehensive segmentation and analysis system, which is initialized with an automatic segmentation model driven by a 2D nnU-Net-based module, precisely delineating muscle area (MA), subcutaneous fat area (SFA), and visceral fat area (VFA). 
The system then provides an interactive interface allowing clinicians to further refine the automated segmentation results with minimal manual effort, ensuring high-quality outcomes. Additionally, the tool automatically computes quantitative parameters for these abdominal tissues, providing clinicians with valuable metrics for patient assessment and treatment planning. 
 
 In summary, our main contributions include:
\begin{itemize}
\item Development of an AI-based tool for automated analysis of abdominal CT body composition, enabling precise localization and segmentation of three abdominal tissue types. This tool automates parameter calculation, streamlining the management of gastrointestinal cancer patients.
\item Introduction of a novel multi-view fusion localization model that accurately extracts the abdominal region from 3D CT scans.
\item Implementation of a high-precision 2D nnU-Net-driven segmentation model for abdominal tissues, complemented by an interactive interface. This combination provides accurate initial results while allowing clinicians to efficiently refine segmentations with minimal effort.
\end{itemize}
\section{Methodology}
\label{sec:format}
\subsection{Overview of the Proposed Automated Tool}
The overview of our AI-driven tool is illustrated in Fig.\ref{fig:overview} and consists of three stages:
\begin{itemize}
    \item Stage I: Abdominal Region localization. Firstly, the user selects a medical image covering the abdomen and feeds it into a multi-view abdomen localization model, aiming to detect the start and end position of abdomen and extract all abdominal slices.
    \item Stage II: Abdominal Tissue Segmentation and Interactive Refinement. Next, all the extracted abdominal slices are individually processed through a well-trained 2D nnUNet-based segmentation model~\cite{nnunet2021} to segment abdominal tissues: muscle, subcutaneous fat area (SFA), and visceral fat area (VFA). Although the segmentation results are already of high quality, users have the option to further optimize these results using an interactive interface based on PyQt5, allowing for the generation of even more refined segmentation labels if needed.
\end{itemize}
\begin{figure*}[htbp]
\centering
\includegraphics[width=0.85\textwidth]{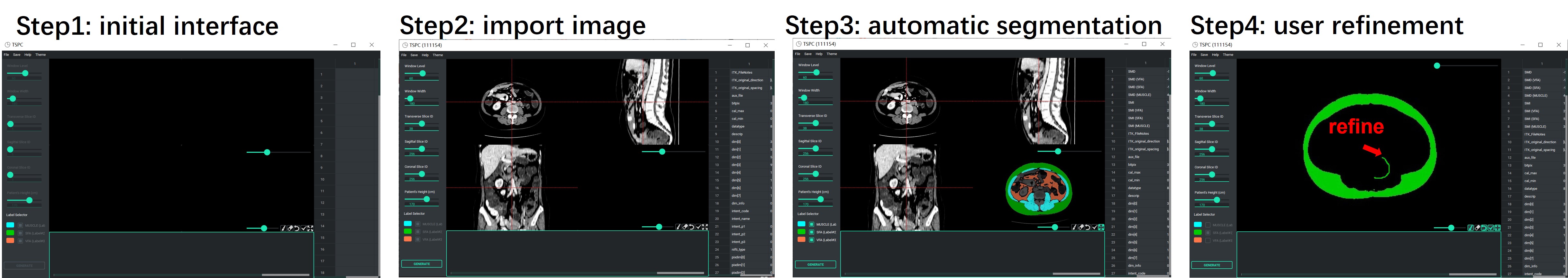}
\caption{Workflow of the interactive automatic annotation tool.}
\label{fig:workflow}
\end{figure*}
\subsection{Step I: Abdominal Region Localization}
The core component of Stage I is the abdominal region localization model. This model begins with preprocessing the 3D CT image through resampling, normalization, and windowing enhancement. The preprocessed image $I_{vol} \in \mathbb {1\times D\times H \times W}$ is then input into a 3D ResNet-18~\cite{resnet3d} to extract volumetric features $F_{vol} \in \mathbb {512 \times D'\times H' \times W'}$ ($D'$=1/16$D$). However, the multiple downsampling stages within the 3D ResNet can cause interference in locating the start and end slice positions, potentially leading to substantial errors in the final predictions. To address this issue, we introduced multi-view CT images to mitigate the loss of positional information during the 3D CT feature extraction process. Inspired by the work of Fahdi et al.~\cite{Kanavati2018AutomaticLS}, we employ a lightweight four-stage 2D downsampling network to extract features from both coronal-view image $I_{cor} \in \mathbb {1\times D\times H}$ and sagittal-view image $I_{sag} \in \mathbb {1\times D\times W}$ centered on the CT volume. Each stage consists of a single convolutional layer, preserving the original positional information of the images as much as possible.
To integrate features from different views, we utilize a multi-view fusion module based on cross-attention, as shown in Eq.\ref{eqa:multiview}. This module comprises two parallel cross-attention modules~\cite{Lin2021CATCA} to generate the volume-coronal fused feature $F_{vol-cor}$ and volume-sagittal fused feature $F_{vol-sag}$. Here, $F_{vol}$ serves as the value and key vectors for both cross-attention modules, while $F_{cor}$ and $F_{sag}$ act as the query vectors for their respective modules.

\begin{align}\label{eqa:multiview}
\widetilde{F}_{vol} &= \nonumber F_{vol-cor} + F_{vol-sag} + F_{vol}\\ 
&= SoftMax\left( \frac{F_{cor}F_{vol}^T}{\sqrt{d_k}} \right)F_{vol}\\
\nonumber&+ SoftMax\left( \frac{F_{sag}F_{vol}^T}{\sqrt{d_k}} \right)F_{vol}
+ F_{vol}
\end{align}
where, $F_{cor}$ and $F_{sag}$ are features extracted from coronal image and sagittal image, respectively. $\widetilde{F}_{vol}$ is the fused feature. The fused feature $\widetilde{F}_{vol}$ is then processed through a linear layer followed by a $SoftMax$ activation function to produce the final 1D Gaussian heatmap predictions of the start and end slice positions of the abdominal region. Inspired by the work of Li.et al.,~\cite{simcc} the probability distribution output vectors are represented as 1D Gaussian heatmaps centered around the predicted start and end positions, respectively. 

\subsection{Stage II: Abdominal Tissue Segmentation and Interactive Refinement}
We extracted abdominal slices based on the start and end positions predicted by the localization model in Stage I. In Stage II, we employ a two-step approach: automatic segmentation followed by interactive refinement.

First, we utilize a 2D segmentation model to perform segmentation predictions on each slice of the volume. We chose nnUNet~\cite{nnunet2021} as our 2D segmentation model due to its exceptional performance in various medical imaging tasks without requiring manual intervention. nnUNet is a deep learning-based segmentation method that automatically configures itself, including pre-processing, network architecture, training, and post-processing for any new task~\cite{nnunet2021}. 
To achieve higher segmentation accuracy and facilitate application to different segmentation tasks, we extracted 2D slices from CT volumetric images to generate a comprehensive 2D abdominal dataset.

This dataset comprises image-mask pairs representing various abdominal components, including muscle, subcutaneous fat area (SFA), and visceral fat area (VFA). 
The 2D nnUNet was then trained on this dataset to predict segmentation masks for these abdominal tissues on each slice.

Following the automatic segmentation, we provide an interactive refinement interface, as illustrated in Fig.\ref{fig:workflow}. This interface allows users, particularly professional doctors, to review and refine the segmentation results generated by the nnUNet model. Users can view the results from transverse, coronal, and sagittal perspectives, enabling a comprehensive assessment of the segmentation quality.

\begin{table*}[h!]
  \begin{center}
    \begin{tabular}{c c c c c c c c c}
      \toprule
      \multicolumn{1}{c}{\multirow{2}*{\textbf{Rep.}}} & \multicolumn{4}{c}{\textbf{start}} & \multicolumn{4}{c}{\textbf{end}}\\
      \cline{2-9}
      &\textbf{Ave. ($mm$)} &\textbf{Max ($mm$)} &\textbf{$\leq 5 mm$} &\textbf{$\leq 10 mm$} &\textbf{Ave. ($mm$)} &\textbf{Max ($mm$)} &\textbf{$\leq 5 mm$} &\textbf{$\leq 10 mm$}\\
      \hline
      \textbf{Gaussian} ($-m$) & 3.08 & 21.97 & 90.0\% & 95.0\% &2.60&12.31& 90.0\% &92.5\%\\
      \textbf{Gaussian} ($-o$) & 3.19 & 27.25 &85.0\% & 92.5\% & 3.53 &34.28&85.0\%&90.0\%\\
      \textbf{0-1} ($-m$) & 5.20 & 17.578 & 57.5\% & 82.5\% & 5.25 & 26.695 & 67.5\% & 82.5\% \\
      \textbf{0-1} ($-o$) & 5.79 & 21.97 & 47.5\% & 87.5\% & 7.62 & 39.06 & 57.5\% & 72.5\% \\
      \bottomrule
    \end{tabular}
    \caption{Evaluation of localization model across various metrics. \textbf{Gaussian} and \textbf{0-1} indicate that the training targets are the Gaussian heatmap and 0-1 distribution, respectively. $-m$ and $-o$ indicate introducing multi-view and one-view, respectively.}\label{tab:tab1}
  \end{center}
\end{table*}
\section{Experiments}
\label{sec:pagestyle}
\subsection{Data Acquisition and pre-processing}
We utilized two datasets from Peking Cancer Hospital to train and test our localization and segmentation models.
The AbdLoc dataset, used for the localization model, consists of 199 CT volumetric images, each annotated with the the starting and ending slice positions of the abdominal region within the volumes. For the segmentation model, we employed the AbdSeg dataset, which encompasses 1,230 image-mask pairs derived from CT images of patients diagnosed with various gastrointestinal cancers. Each slice in the dataset was extracted from abdominal region and is accompanied by a non-empty mask. For the localization model, we preprocessed the AbdLoc data by resampling each volume to a fixed size of $512\times 64\times 64$ before feeding it into the model. For the segmentation model, we extracted multiple 2D slices from each 3D volume in the AbdSeg dataset, resulting in a total of 85,944 2D image-mask pairs. The preprocessing of these 2D images followed the automated pipeline established by nnUNet~\cite{nnunet2021}.
\subsection{Implementation Details}
For the localization model, we implemented a 3D ResNet-18 for feature extraction from volumetric input, complemented by a four-stage lightweight downsampling network for coronal and sagittal inputs. We divided the AbdLoc dataset into training and testing sets in a 4:1 ratio. The model was trained for 250 epochs using the Adam optimizer with an initial learning rate of $1e-3$.
We employed Kullback-Leibler divergence as the loss function, calculated between the predicted and ground truth 1D Gaussian heatmaps. Experiments for both training and testing of the localization model were conducted on two NVIDIA Tesla A800 GPUs, with a batch size of 8 for each GPU.
For the segmentation model, we trained a 2D nnUNet using 68,756 samples for training and 17,188 samples for validation. nnU-Net automatically configures the entire segmentation pipeline, encompassing all steps from preprocessing to model configuration, model training and post-processing. The experiments for the segmentation model were conducted on a single NIVIDA Tesla A800 GPU.
\subsection{Evaluation Metrics}
For the localization model, we employed the $L1$ distance metric as the primary evaluation criterion. Our process began with resampling all input images to a uniform size of $512\times 64\times 64$, recording the spacing values both before ($s_{ori}$) and after ($s_{res}$) resampling. Using these spacing values, along with the predicted ($pred$) and ground-truth ($gt$) slice indexes, we calculated the prediction error for the start and end positions, denoted as $L_{err}(mm)$. The $L1$ error between ground truth and prediction is calculated as follows:
\begin{equation}
\begin{aligned}
    L_{err} = \Vert pred\times s_{res}-gt\times s_{ori}\Vert 
\end{aligned}
\end{equation}
Additionally, we used the Dice Score Coefficient (DSC), the 95th percentile Hausdorff Distance (95th HD) and Intersection over Union (IoU) to evaluate segmentation quality on the testing dataset.

\subsection{Localization and Segmentation Results Analysis}
We evaluated the accuracy of the localization model using 40 CT volumetric samples. The assessment involved four metrics: the average $L1$ error, the maximum $L1$ error, the proportion of $L1$ error $\leq 5\ mm$, and the proportion of $L1$ error $\leq 10\ mm$. An $L1\leq5 mm$ can be considered indicative of a correct localization prediction.

To validate the efficacy of representing the start and end slice positions with Gaussian heatmaps, we also conducted experiments using 0-1 distribution(where 1 indicates the start or end index position), as shown in Tab.\ref{tab:tab1}. The results clearly demonstrate that the Gaussian heatmap-based localization model achieved superior accuracy, outperforming the 0-1 distribution-based model across nearly all evaluated metrics.
Furthermore, we performed ablation experiments presented in Tab.\ref{tab:tab1}, which indicate that the introduction of multi-view approaches effectively enhances localization performance, as evidenced by the reduction in both average and maximum localization errors.

For the segmentation performance, we assessed 17,188 2D CT scans using the DSC, 95th HD and IoU. Tab.\ref{tab:tab2} presents the evaluation results for segmenting muscle, SFA, and VFA. The average DSC of the three tissue types exceeds 0.95, indicating highly reliable segmentation predictions with only slight discrepancies between the predicted masks and the ground truth masks. The average 95th HD of approximately $2 mm$ further confirms the high precision of our segmentation results. The average IoU of the three tissue types also exceeds 90\%, exhibiting highly reliable segmentation predictions as well.


\begin{table}[h!]
  \begin{center}
    \begin{tabular}{c c c c c}
    \toprule
       & muscle & SFA & VFA & Ave.\\
      \hline
      \textbf{DSC} &0.979 & 0.979 &0.944 &0.967 \\
      \textbf{95 HD} &1.179 &1.427 &3.612 &2.073\\
      \textbf{IoU} &95.95\% &94.98\% &88.84\% &93.26\%\\
      \bottomrule
    \end{tabular}
    \caption{Evaluation of the segmentation model for muscle, SFA and VFA according to DSC, 95 HD and IoU.}\label{tab:tab2}
  \end{center}
\end{table}
\section{Conclusion}
\label{sec:print}
This study presents a novel AI-driven tool for automated analysis of abdominal CT body composition, specifically designed to assist in the management of gastrointestinal cancer patients. By integrating a multi-view abdominal localization model and a high-precision segmentation model based on a 2D nnU-Net, the proposed tool offers promising improvements in efficiency and accuracy for abdominal tissue analysis. Moreover, the interactive interface enables clinicians to refine segmentation results with minimal manual effort, further enhancing the quality of analysis. This combination of automation, precision, and collaborative features has the potential to standardize abdominal tissue measurement, reduce clinician workload, and ultimately improve the diagnosis and treatment of gastrointestinal cancer patients.
At present, this work only focuses on the gastric cancer and localization and segmentation of abdominal cavity, and future work can be extended to various types of cancer and organ locations.
\clearpage
\section{Compliance with ethical standards}
This study was approved by the Institutional Review Board of Beijing Cancer Hospital(2024KT74), and was conducted according to the tenets of the Declaration of Helsinki. Written informed consent was revoked for this retrospective study by IRB of Beijing Cancer Hospital.



\section{Acknowledgments}
This work was supported by the National Natural Science Foundation of China(12090022 to B.D., 11831002 to B.D., 81801778 to L.Z.), Clinical Medicine Plus X-Young Scholars Project of Peking University (PKU2023LCXQ041 to L.Z.).
\bibliographystyle{ieeetr}
\bibliography{refs}

\end{document}